\documentclass{aastex63}

\usepackage{amsmath,amstext}
\usepackage[T1]{fontenc}
\usepackage[figure,figure*]{hypcap}
\usepackage{float}
\usepackage{wasysym}
\usepackage[normalem]{ulem}
\usepackage{enumitem,kantlipsum}

\usepackage{booktabs,multirow,array}
\submitjournal{ApJ}

\shorttitle{Probing Reionization with the VPF}
\graphicspath{{./}{figures/}}

\begin{document}

\title{Probing Patchy Reionization with the Void Probability Function of Lyman-$\alpha$ Emitters}

\correspondingauthor{Lucia A. Perez}
\email{lucia.perez.phd@gmail.com}

\author[0000-0002-8449-1956]{Lucia A. Perez}
\affiliation{Department of Astrophysical Sciences,
Princeton University,
4 Ivy Lane,
Princeton, NJ 08544, USA}
\affiliation{Arizona State University,
School of Earth and Space Exploration,
781 Terrace Mall,
Tempe, AZ 85287, USA}

\author[0000-0002-9226-5350]{Sangeeta Malhotra}
\affiliation{NASA Goddard Space Flight Center,
8800 Greenbelt Road,
Greenbelt, MD 20771, USA}

\author[0000-0002-1501-454X]{James E. Rhoads}
\affiliation{NASA Goddard Space Flight Center,
8800 Greenbelt Road,
Greenbelt, MD 20771, USA}

\author[0000-0003-4207-0245]{Peter Laursen}
\affiliation{Cosmic Dawn Center (DAWN) and Niels Bohr Institute,
University of Copenhagen,
Jagtvej 128, 2200 Copenhagen N, Denmark}

\author[0000-0002-0784-1852]{Isak G.B. Wold}
\affiliation{NASA Goddard Space Flight Center,
8800 Greenbelt Road,
Greenbelt, MD 20771, USA}
\affil{Department of Physics, The Catholic University of America, Washington, DC 20064, USA }
\affil{Center for Research and Exploration in Space Science and Technology, NASA/GSFC, Greenbelt, MD 20771, USA}

\begin{abstract}
We probe what constraints for the global ionized hydrogen fraction the Void Probability Function (VPF) clustering can give for the Lyman-Alpha Galaxies in the Epoch of Reionization (LAGER) narrowband survey as a function of area. Neutral hydrogen acts like a fog for Lyman-alpha emission, and measuring the drop in the luminosity function of Lyman-$\alpha$ emitters (LAEs) has been used to constrain the ionization fraction in narrowband surveys. However, the clustering of LAEs is independent from the luminosity function's inherent evolution, and can offer additional constraints for reionization under different models. The VPF measures how likely a given circle is to be empty.
It is a volume-averaged clustering statistic that traces the behavior of higher order correlations, and its simplicity offers helpful frameworks for planning surveys.

Using the \citet{Jensen2014} simulations of LAEs within various amount of ionized intergalactic medium, we predict the behavior of the VPF in one (301x150.5x30 Mpc$^3$), four (5.44$\times 10^6$ Mpc$^3$), or eight (1.1$\times 10^7$ Mpc$^3$) fields of LAGER imaging. 
We examine the VPF at 5 and 13 arcminutes, corresponding to the minimum scale implied by the LAE density and the separation of the 2D VPF from random, and the maximum scale from the 8-field 15.5 deg$^2$ LAGER area.
We find that even a single DECam field of LAGER (2-3 deg$^2$) could  discriminate between mostly neutral vs. ionized. Additionally, we find four fields allows the distinction between 30, 50, and 95 percent ionized; and that eight fields could even distinguish between 30, 50, 73, and 95 percent ionized. 
\end{abstract}


\section{Introduction} \label{sec:intro}

Lyman-$\alpha$ emitters (LAEs) are unique galaxies whose properties make them excellent probes of various phenomena at high redshift. High-redshift LAEs can be selected using narrowband imaging in the optical and infrared, and are particularly suited to study the epoch of reionization (EoR). The EoR was a key era and final phase change of the universe where the opaque `fog' of early neutral hydrogen was ionized by the earliest galaxies. By leveraging the strong emission in the resonant Ly-$\alpha$ line, the observed properties and distribution of $z>6$ LAEs can be used to understand the extent of neutral hydrogen around them and the process and pacing of reionization (\citealt{MiraldaEscude1998,RhoadsMalhotra2001,EMHu2002,Furlanetto2006,Mesinger+Furlanetto2008, McQuinn2007}). 

LAEs sensitively indicate the presence of neutral hydrogen around them as it attenuates their brightness and luminosity function (e.g. \citealt{HaimanSpaans1999,Santos2004, Furlanetto2004}), and the extent of ionized hydrogen as it reveals clusters of LAEs (e.g. \citealt{Hu2021, Sobral2015}; and others). Much work has been done to constrain and understand reionization using the evolution of the Ly$\alpha$ luminosity function (LF) across redshift history (e.g. \citealt{Ajiki2003, Ouchi2003, Hu2004, MalhotraRhoads2004, Ouchi2010, Santos2016, Ouchi2018, Morales2021}). \citet{McQuinn2007} modeled the expected suppression of the Ly$\alpha$ LF across many neutral fractions for the global IGM, contributing to constraints of neutral hydrogen fraction to between 0.2-0.4 at $z\sim7$ (seen in other LAE surveys, e.g. \citealt{Konno2018}). Similar modeling from \citet{Santos2004} contributed to similar constraints in \citet{Ouchi2010}. Some works have found no evidence of attenuation in the $z>6$ Ly$\alpha$ LF in their analysis (\citealt{MalhotraRhoads2004,Tilvi2010}); while some analyses have measured an attenuation in the LFs or LAE number densities within $5.7<z<7$ (\citealt{Kashikawa2006, Iye2006, Ouchi2010, Kashikawa2011, Konno2014, Konno2018}). 

The Lyman-Alpha Galaxies in the Epoch of Reionization (LAGER) survey is the largest narrowband survey yet of LAEs during reionization, on track to cover $\sim$24 deg$^2$ and better constrain the timing and morphology of reionization. The survey utilizes a custom-made N964 narrowband filter centered at 9642{\AA} with a filter FWHM of 92{\AA} \citep{Zheng2019} that exploits the 3 deg$^2$ field of view of the Dark Energy Camera (DECam) instrument mounted on the Blanco 4m telescope at Cerro Tololo International Observatory. LAGER has to date observed 195 z=6.9 LAEs across four fields and 10.19 deg$^2$ (\citealt{Zheng2017,Hu2019, Wold2022}), making it the largest survey of LAEs near z$\sim$7 to date with four additional fields still to be fully observed and analyzed. With a high rate of spectroscopic confirmation of LAEs (\citealt{Hu2017, Yang2019}; Harish et al. submitted) and a survey volume at least twice as large as others used in Ly$\alpha$ LF analyses \citep{Wold2022}, LAGER is a powerful and efficient survey of Ly$\alpha$ during reionization. So far, the LAGER Ly$\alpha$ luminosity functions and their evolution are consistent with a nearly completely ionized universe at z=6.9 (\citealt{Hu2019, Wold2022}).

Neutral fraction constraints from studies of the Ly$\alpha$ LF can show tension with the measurements of the IGM temperature (which indicate a mostly ionized medium at $z\sim6$, e.g. \citealt{Fan2006}) and measurements of the Ly$\alpha$ line profile (e.g. \citealt{Ouchi2010} rule out a fully neutral universe at $z=6.6$ with the model of \citealt{HaimanCen2005}). However, the Ly$\alpha$ LF analyses come with the caveat that the LF suppression might be explained by the evolution of the halo mass function \citep{Dijkstra2007}, cosmic variance, details of the chosen model, or changes in when galaxies show strong Ly$\alpha$ emission (e.g. \citealt{Ota2008, Stark2010, Pentericci2011, Ono2012, Schenker2012, Endsley2021}). A way to break these tensions and degeneracy is with the clustering of LAEs during the epoch of reionization, which offers an additional way to constrain reionization that can circumvent possible galaxy evolution \citep{McQuinn2007} and is independent from the evolution of the intrinsic LF.


In this work, we focus on what constraints for reionization the Void Probability Function (VPF) might give for LAEs selected from large-volume narrowband surveys.   Such surveys typically cover large areas with a redshift extent limited by the filter bandpass.  This results in a survey geometry that is wider but with smaller $\Delta z$ than is typical for surveys that begin with selection by broad-band dropout, photometric redshift, or direct spectroscopic searches.  We base our expectations for survey areas and sensitivities on LAGER, but note that our results may be applied more generally.   Specifically, we seek to quantify what precision on the constraints of reionization LAGER or similar narrowband surveys may find with clustering, as a function of imaged area.

The VPF is a statistical measure of clustering that simply asks: how likely is a circle or sphere of a given size to be empty in the sample? It is also the zero-point volume-averaged correlation function, and is measured by counting how many cells of a given size are empty, often alongside count-in-cells. As the 0$^{\text{th}}$ moment of count-in-cells, it carries the signature of higher-order correlation functions beyond two-points, and its simplicity can be leveraged to guide the number density and volume of surveys \citep{Perez2021}. \citet{Kashikawa2006}, \citet{McQuinn2007}, and \citet{Gangolli2021} have examined the ability of the VPF to constrain reionization for various generations of Subaru LAE observations at $z=6.6$ and $z=5.7$. Here, we focus on the constraints and survey guidelines that the VPF can give for the uniquely large and growing LAGER survey of LAEs at $z=6.9$ in tandem with the LAGER luminosity function analysis. 

This paper is organized as follows. In \textsection \ref{sec:SimsMethods}, we describe our use of the \citet{Jensen2014} simulations of LAEs within different IGM fractions of neutral hydrogen to predict the VPF of LAGER LAEs. In \textsection \ref{sec:EoRClustering_General}, we introduce the use of clustering to constrain the fraction of neutral hydrogen during the epoch of reionization, and motivate the focus on the Void Probability Function for the LAGER narrowband survey. In \textsection \ref{sec:GetXiFractionfromVPF}, we ask and answer: how distinguishable are the different ionization fractions using the VPF for narrowband-detected LAGER LAEs, as a function of survey area? In particular, we leverage the large simulation volume at various ionization fractions to measure the VPF in mock LAE slices that mimic a single LAGER field, the four currently imaged, and the eight in the full survey plan. Our work and conclusions are summarized in \textsection \ref{sec:Conclusion}.

\section{Simulations and Methodology}\label{sec:SimsMethods}

\subsection{The Jensen et al. (2014) simulations of LAEs during reionization}  \label{subsec:JensenSimDeets}


To obtain the ionization fraction and the Ly$\alpha$ luminosities of galaxies, we make use of the simulations of \citealt{Jensen2014} who modeled LAEs during the epoch of reionization (expanding upon \citealt{Jensen2013}). The simulations exist at the specific mass-averaged ionized hydrogen fractions in the IGM of $\langle x_i \rangle_m=\{0.30, 0.50, 0.58, 0.73, 0.83, 0.92, 0.95\}$, and for our applications, give each LAE's transmitted Ly$\alpha$ luminosity. The strength of this model is its combination of a large volume, ensuring a statistically sound sample of galaxies, and high-resolution radiative transfer. For example, we find the cosmic variance across the entire \citet{Jensen2014} volumes is less than 4$\%$ after all selections using the \citet{TrentiStiavelli2008} cosmic variance calculator \footnote{\url{https://www.ph.unimelb.edu.au/~mtrenti/cvc/CosmicVariance.html}}. Here we briefly summarize the model, and direct readers to those works for full details.

The large-scale structure of the Universe was modeled using a 165 billion particle $N$-body simulation of a (602x607x600) Mpc$^3$ volume. Halos were populated with galaxies to match the UV Lyman-break and Ly$\alpha$ luminosity functions of \citet{Ouchi2010}. Each galaxy was modelled as emitting an intrinsic, halo mass-dependent, double-peaked Ly$\alpha$ spectrum. 
The intrinsic Ly$\alpha$ luminosity that emerges from a given LAE at 1.5$r_{vir}$ was randomly drawn from a lognormal distribution with a standard deviation of $\sigma$ = 0.4 dex and a mean proportional to the host dark matter halo's mass. Details of the "Gaussian-minus-Gaussian" line shape recipe used to generate the double-peaked spectra can be found in \citet{Jensen2013}. Subsequently, the individual Ly$\alpha$ spectra were modified to account for scattering in the IGM, using the radiative transfer (RT) code {\sc IGMtransfer} \citep{IGMtransfer}. The RT code calculates the transmission across the Ly$\alpha$ line through the circum- and intergalactic medium, on the basis of the much higher-resolution, hydrodynamic, cosmological simulations of \citet{Laursen2011}. The wavelength-dependent transmission is defined as the median value of sightlines in all directions, of all galaxies in the simulation. Sightlines began at a distance of 1.5 virial radii from the center of a given galaxy, at which distance most of the Ly$\alpha$ photons have experienced their last scattering \emph{into} the line of sight, after which they are only scattered \emph{out of} the line of sight \citep{Laursen2011}. The resulting direction-dependent transmitted Ly$\alpha$ luminosities are the selection criteria for our LAEs.



Next, we highlight these and other assumptions that \citet{Jensen2014} made in their single simulated reionization history. When simulating the radiative transfer of ionizing radiation out of galaxies, each source from the N-body simulation was given a flux proportional to its mass. Therefore, the photoionization rate of each source (Eq. 1 in \citealt{Jensen2013}) depends on assumptions for the initial mass function, star formation efficiency, and escape fraction of the galaxies. Additionally, \citet{Jensen2013} assume that only galaxies took part in reionization, and that small sources with $M_h < 10^9 M_{\odot}$ were suppressed once the IGM around them was more than 10$\%$ ionized (a.k.a. self-regulation). This model therefore generated \textit{inside-out reionization}, where the first regions to be ionized are those that have the most galaxies and are the highest mass (confirmed by the higher mass-averaged ionization fractions than volume-averaged at a given redshift). The \citet{Jensen2013} and \citet{Jensen2014} simulations did not include the effects of gas clumping and Lyman limit systems, and give the disclaimer that they may have overestimated the redshift when reionization ended. 
They also assumed that the true topology of reionization would be consistent with their model (citing \citealt{Friedrich2011} and \citealt{Iliev2012}), and therefore analyzed and shared their simulations with \textit{mass-averaged} ionization fractions rather than volume-averaged. 
The mass-averaged ionization fractions correspond approximately to volume-averaged ionized hydrogen fractions of $\langle x_i \rangle_v=\{0.22, 0.40, 0.485, 0.66, 0.78, 0.89, 0.93\}$, according to Fig. 2 of \citet{Jensen2013}.

Various observations were used to fine tune and calibrate the assumptions made when creating these simulations. To study the volume at a particular ionization fraction, they selected whatever redshift corresponds to the sought fraction and scale the dark matter halo masses so that the intrinsic luminosity function of the volume still matched the observed Ly$\alpha$ LF at z = 5.7 \citep{Ouchi2010}. The simulation is by default scaled to $z=6.5$, which allows our analysis for LAGER at $z=6.9$. The rest equivalent width (EW) distribution followed a lognormal EW distribution supported by lower redshift observations (e.g. \citealt{ReddySteidel2009}, as it was not well-known at z=6.5). Galaxies are randomly assigned EWs with no correlation to Ly$\alpha$ luminosity. The EW distribution and assignment was found to be consistent with (\citealt{Stark2010, Jiang2013, Hayes2014, Zheng2014}). \citet{Jensen2013} and \citet{Jensen2014} took the z=5.7 Ly$\alpha$ luminosity function of \citet{Ouchi2010} as true, and match the mass-to-light ratio of their luminosity to it.

With regards to LAE clustering, \citet{Jensen2013} first ran smaller simulations with this model, and compared their analyses of the simulated LAEs to: the observed Ly$\alpha$ luminosity functions of \citet{Ouchi2010} and \citet{Kashikawa2011}; measured decreases in the observed LAE fraction; and the 3D and 2D two-point correlation functions of observed LAEs (measured $\xi(r)$ and its projection into $w(\theta)$; \citealt{Ouchi2010}), all as a function of $\langle x_i \rangle _m$. With expanded simulations, \citet{Jensen2014} showed that CiC could tell apart ionization fractions 
and probe the unique signal of inhomogeneous reionization on the clustering of LAEs \citep{McQuinn2007}. We add the VPF to the repertoire of clustering statistics used on these simulations, alongside the rest of the work we present. 


Throughout this work, we use transmitted Ly$\alpha$ luminosity in the z-direction in any place we refer to the luminosities of the simulated LAEs. We use the \textit{astropy}\footnote{\url{https://www.astropy.org}: a community-developed core Python package for astronomy \citep{astropy:2013, astropy:2018}.} cosmology package under a Planck 2013 flat $\Lambda$CDM cosmology, with $\Omega_{M,0}$=0.307, $\Omega_{b,0}$=0.0483, $H_0$=67.8 km (Mpc s)$^{-1}$. The original simulations were run with a flat $\Lambda$ cold dark matter consistent with the 9 year Wilkinson Microwave Anisotropy Probe model with ($\Omega_m$, $\Omega_b$, h, n, $\sigma_8$) = (0.27, 0.044, 0.7, 0.96, 0.8), WMAP results \citep{WMAPcosmo}, and we note this difference in cosmology creates a negligible $>0.5\%$ change in the surface densities and radii we list. 

\begin{table*}[t]
	\begin{center}
    \caption{Details of our use of the \citet{Jensen2014} simulations of LAEs in reionization for the LAGER VPF tests. The mass-averaged ionized hydrogen fraction in the IGM of each simulation is $\langle x_i\rangle _m$; the corresponding volume-averaged ionized hydrogen fraction in the IGM (from Figure 2 of \citealt{Jensen2013}) is $\langle x_i\rangle _v$; N$_{\text{total}}$ is the total number of LAEs in each simulation before any selection; and N$_{\text{selected}}$ is the number of LAEs after the LAGER luminosity cut of log$_{10}$(L$_{\text{cut}}$)=\textbf{42.9} erg s$^{-1}$. \citet{Wold2022} conservatively conclude, based on their analysis of the 4-field LAGER luminosity function with 195 LAEs, that at z=6.9 the neutral hydrogen fraction $\langle x_{\mathrm{HI}} \rangle_m$ < 0.33, or in the formalism of the \citet{Jensen2014} simulations, $\langle x_{i,\mathrm{LAGER}} \rangle_m$ > 0.67. 
    The volume density is calculated with the entire (602x607x600) Mpc$^3$ simulation, and the surface density is averaged across the 20 narrowband-thin `slices' of (602x607x30) Mpc$^3$.
    \label{table:LAGERSimInfoTable}
    }   
	\begin{tabular}{p{1.0cm}p{1.25cm}p{1.5cm}p{2cm}p{3.5cm}p{3.5cm}} 
	\hline \hline
    $\langle x_i\rangle _m$ & $\langle x_i\rangle _v$ & N$_{\text{total}}$ & N$_{\text{selected}}$ & Volume Density $\mathcal{N}_{\text{Mpc}^{-3}}$ & Surface Density $\Sigma_{\text{amin}^{-2}}$  \\
    \hline
        4-field & LAGER & ... & 195 & 2.86 $\times 10^{-5}$ & 4.74 $\times 10^{-3}$ \\
        \hline
		0.30 & 0.22 & 978,687 & 2,765 & 1.27 $\times 10^{-5}$ & 2.47 $\times 10^{-3}$ \\
        0.50 & 0.40 & 1,632,805 & 3,870 & 1.78 $\times 10^{-5}$ & 3.46 $\times 10^{-3}$ \\
        0.73 & 0.66 & 2,301,037 & 6,725 & 3.09 $\times 10^{-5}$ & 6.01 $\times 10^{-3}$ \\
        0.95 & 0.93 & 3,052,695 & 7,124 & 3.28 $\times 10^{-5}$ & 6.37 $\times 10^{-3}$ \\
    \hline \hline	
	\end{tabular}
	\end{center}
    
\end{table*}

\begin{figure*}
	\begin{center}
\includegraphics[width=.8\textwidth]{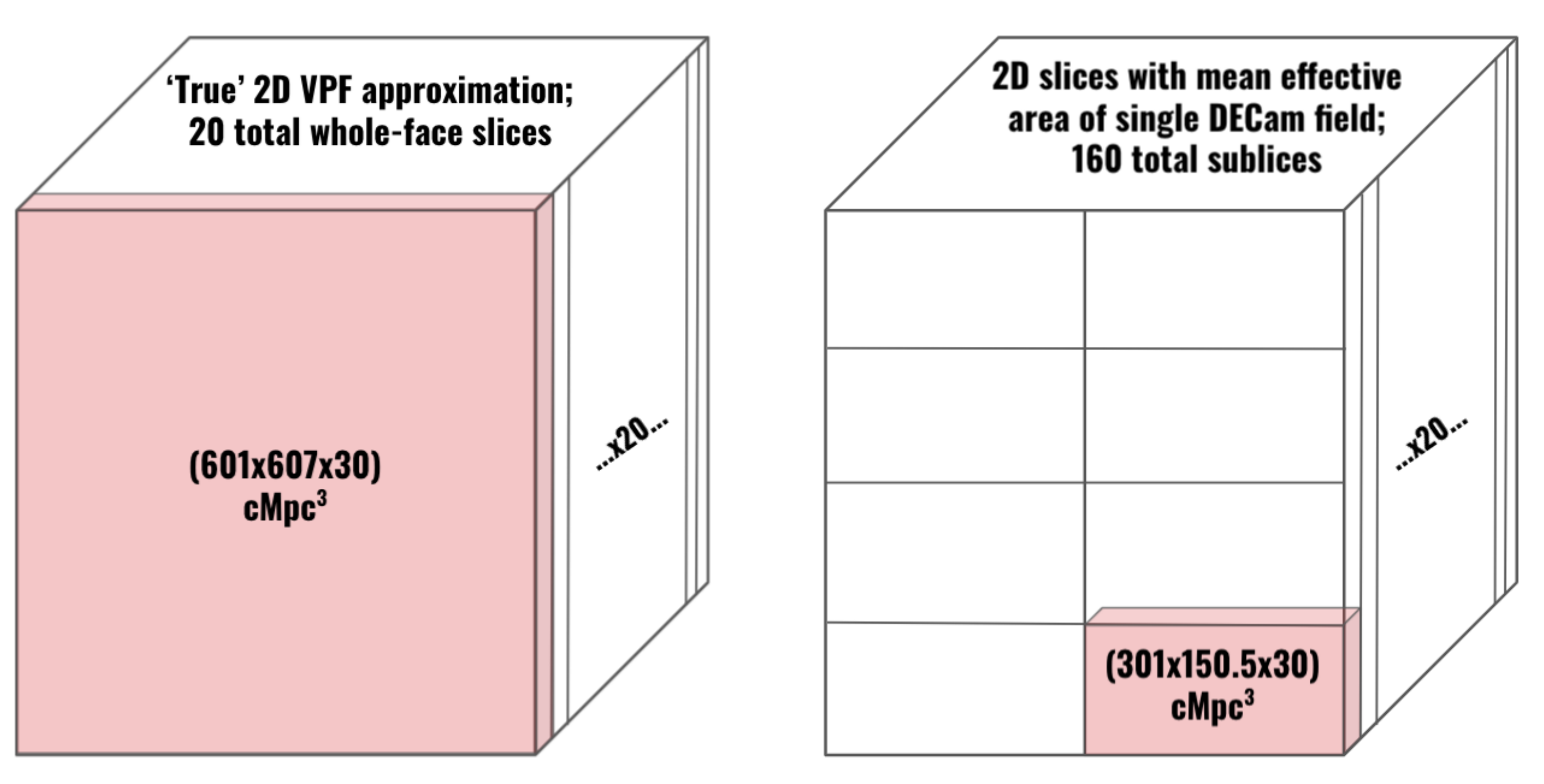}
	\caption{A schematic showing how we slice the \cite{Jensen2014} simulations in order to
	explore how the precision of reionization constraints changes with area using the VPF in LAGER.
	To approximate the `true' 2D VPF value for the LAEs, we create 20 full-face slices of (601x607x30 cMpc)$^3$. This area corresponds to what the 8 currently planned fields of LAGER would likely cover for z=6.9 LAEs, were they all connected. This approach allows us to average over cosmic variance within the scale of the simulations, as well as include the signal of any large voids what might be split by the RA and Dec partitions we chose. To test what LAGER will measure in individual fields and surveys of different sizes, we create 160 subslices of (300x150.5x30 cMpc)$^3$. This corresponds to the approximate mean effective area across a single LAGER field ($\sim$ 2 deg$^2$), as observed and analyzed in \citet{Wold2022}. In Sections \ref{sec:GetXiFractionfromVPF}, we randomly sample one, four, or eight  subslices to predict the behavior and reionization constraints of the VPF once it is measured in the LAGER LAEs.
    \label{fig:SlicesDiagram} }
	
	\end{center}
\end{figure*}

\subsection{Creating mock LAGER LAE volumes for VPF predictions}\label{subsec:LAGERsetup}

As the LAGER collaboration prepares for the clustering analysis of our LAEs and expands observations to more fields, we consider: what additional and complementary constraints can the VPF give for the global ionization fraction during the epoch of reionization? In this section, we parse up the ($\sim$ 600 cMpc)$^3$ volume of \citet{Jensen2014} into subslices that mimic the breadth of the LAGER at $z=6.9$, to prepare to create predictions for the 2D VPF clustering of the narrowband-selected LAEs of LAGER. 

First, we tune our LAE selection to match what LAGER has observed in the published 195 $z=6.9$ LAEs utilized in the 4-field LAGER luminosity function \citep{Wold2022} with an 
average 50$\%$ complete sample given a luminosity limit of log$_{10}$L$_{\text{Ly}\alpha}>$ 42.9 erg s$^{-1}$ and equivalent width (EW) threshold of EW$_{\text{Ly}\alpha} \gtrsim$ 10\AA. For such narrow imaging (approximately 30 cMpc for the LAGER narrowband), the flux-limited observations of LAEs mimics a line luminosity cut that would create a volume-limited sample. 
LAEs are known to have a possible duty cycle, meaning not all galaxies capable of emitting Ly$\alpha$ will be observed in the line. We therefore apply a duty cycle of 30$\%$ to the \citet{Jensen2014} simulations after a luminosity cut of log$_{10}$L$_{\text{Ly}\alpha}>$ 42.9 erg s$^{-1}$, following \citet{Kovac2007}, and find excellent agreement between the observed LAE number density in the current 4 fields of LAGER \citep{Wold2022} and the $\langle x_i\rangle _m=0.73$ simulation (closest to the derived limit of $\langle x_{i, \textrm{LAGER}} \rangle > 0.67$). 

We also confirm the simulations show great consistency with the observed number and surface densities of LAEs across the current 4 fields of LAGER, $\mathcal{N}_{\text{LAGER}}= 2.86 \times 10^{-5}$ Mpc$^{-3}$ and $\Sigma_{\text{LAGER}} = 4.74 \times 10^{-3}$ arcmin$^{-2}$. The number densities of the simulated LAEs range from $1.12\times 10^{-5}$ (30$\%$ ionized) to $3.11\times 10^{-5}$ (95$\%$ ionized) Mpc$^{-3}$. Table \ref{table:LAGERSimInfoTable} details the number and densities of LAEs in the different simulations we use after the our selection. Finally, in this work we apply a selection to mimic the LAGER sample, easing comparisons to simulations and other surveys. However, we confirm in the Appendix that applying the complete and detailed LAGER incompleteness analysis in \citet[e.g.\ Fig.\ 6]{Wold2022} yields nearly identical conclusions. We show the VPF measurements of the \citet{Jensen2014} LAE catalogs using the detailed observed LAGER incompleteness analysis in the Appendix.

We now briefly summarize the observations LAGER has already completed and analyzed to motivate our slicing and sampling of the \citet{Jensen2014} simulations. \citet{Wold2022} detect their $z=6.9$ LAEs in a total area of 10.19 deg$^2$ across the WIDE12 (3.24 deg$^2$), GAMA15A (2.91 deg$^2$), COSMOS (1.90 deg$^2$), and CDFS (2.14 deg$^2$) fields, or a total effective area of 7.6 deg$^2$. At $z=6.9$, a cube face of (601x607 cMpc)$^2$ of the \citet{Jensen2014} simulations covers an area of 15.5 deg$^2$, allowing us to predict what the full 8 fields of LAGER may image if the same effective areas stand for future fields. We clarify here that although each DECam image observes 3.3 deg$^2$, each field in the \citet{Wold2022} LAE $z=6.9$ luminosity function analysis averages 2.55 deg$^2$ due to a combination of bright foreground stars, limited ancillary data for LAE selection, and gaps from not dithering the detector in very early stages of the LAGER observations. We 
assume the future LAGER fields will cover at least similar effective areas, corresponding to the (301x150.5x30) cMpc$^3$ or $\sim$2 deg$^2$ slices. Finally, we emphasize that this work with the VPF and the \citet{Jensen2014} simulations is an independent assessment of the precision LAGER may find in its reionization constraints using clustering.
This work was done independently to the work of \citet{Wold2022}, had mostly taken shape before their work was completed, and is at its core agnostic to their results. We include their results here for context, observational grounding for part of the experiment set up, and to demonstrate that the predicted area coverage of LAGER was indeed met by the team's observations.

Figure \ref{fig:SlicesDiagram} illustrates how we slice the \citet{Jensen2014} simulations for these LAGER tests. To get a sense of what the `true' 2D VPF would be for the LAEs at this redshift and under the given ionization fraction, we create 20 slices of (602x607x30 cMpc)$^3$. To test what ionization fractions one, four, or eight sampled LAGER fields might probe with the VPF, we create 160 independent slices of (300x150x30) Mpc$^3$, corresponding to the average effective area of one LAGER imaging according to the 4 field analysis of \citet{Wold2022}, and the approximate distance covered by the N964 filter. 
We test what constraints the VPF of LAGER will be able to give for reionization for a single DECam field, the 4 fields that have been imaged and processed as of the writing of this work, and the total 8 fields planned for LAGER. Our choice of slices also incorporates the effects of cosmic variance on the VPF: in applying the cosmic variance calculator of \citet{TrentiStiavelli2008} to the individual DECam-field-sized slices, we find a total fractional error of $\sim24-30\%$, which is encompassed in the range of LAE counts we measure in the slices.

\section{Constraining reionization with the Void Probability Function}\label{sec:EoRClustering_General}

\subsection{The Void Probability Function as a probe}\label{subsec:VPFnEoR}

Neutral hydrogen acts like a fog for Lyman-$\alpha$ emission, meaning that LAEs during the epoch of reionization are influenced by the amount and distribution of neutral hydrogen in the nearby IGM. The effect of reionization on LAEs can be subtle and measuring it quite model-dependent, so constraining this effect in various ways allows for stronger scientific consensus. Different clustering tools probe different aspects of galaxies' large scale structure and vary in performance at different scales. Several studies have been done with the angular correlation function (ACF) of narrowband-selected LAEs during reionization (\citealt{Ouchi2010, SobbachiMesinger2015, Ouchi2018, Gangolli2021}) to constrain reionization with LAEs. However, small samples sizes, the resulting Poisson noise in the ACF, and the requirement of several distance scales to derive a correlation length can lead to significant levels of uncertainty in this vital constraint on reionization. 

As the field gathers and awaits larger samples of observed LAEs during reionization to more finely measure the clustering signals caused by inhomogeneous reionization, these effects have been mostly explored in simulated LAEs. \citet{Jensen2013} explored the two-point correlation function in their early simulations, and later showed that count-in-cells showed a difference between ionization fractions under a single number density cut in \citet{Jensen2014}. Recently, \citet{Gangolli2021} compared how several statistics, including the two-dimensional VPF, are able to constrain late reionization models and probe the effect of cosmic variance in mock LAE surveys made to mimic a SILVERRUSH-like survey at $z=5.7$ and $6.6$ (\citealt{Ouchi2018, Konno2018}). They found that the VPF is more sensitive than the angular correlation function when testing very late-stage ionization models under a simple $\chi^2$ analysis, even when incorporating high contamination fractions. 

The Void Probability Function (VPF) can complement the ACF analysis by analyzing the volume-averaged clustering instead, and the guidelines of \citet{Perez2021} can help refine the scales of surveys and their analyses, as well as probe uncertainties in different ways. The VPF measures the probability of a given region being devoid of galaxies. It can also be thought of as a volume-averaged zero-point correlation function, and can connect to higher order correlation functions under hierarchical scaling frameworks. The VPF often complements count-in-cells (CiC) analyses, especially those that study the underlying physics behind hierarchical scaling (e.g. \citealt{Conroy2005}). Like CiC, the VPF is known theoretically for a given number density $\mathcal{N}$: $P_N = ( (\mathcal{N}V)^{N}/N!) \exp(-\mathcal{N}V)$, where $N=0$ for the VPF so $P_0=\text{VPF}=\exp(-\mathcal{N}V)$. The VPF can also give meaningful results at few distance scales, rather than the several the correlation function requires to confidently measure a power law. 

The VPF is faster to calculate per-capita than CiC, as it only focuses on cells with zero galaxies, and as is motivated in \citet{Perez2021}, has more intuitive errors and limits than CiC. \citet{Perez2021} also contend that for applications that do not require a correlation length (with the standard two-point correlation function) or an analysis of hierarchical scaling (with count-in-cells)--such as detecting an excess of LAE clustering due to inhomogeneous reionization--the speedy VPF and its intuitive bounds and errors might be a preferable option. In this work predicting what constraints for reionization LAGER may give with the VPF, we use both self-written algorithms developed in \citet{Perez2021} as well as the incredibly fast $k$-nearest neighbor method introduced in \citet{Banerjee2020} to measure the VPF.

\subsection{When can we trust a VPF measurement?}\label{subsec:WhenuseVPF}

A key goal of this work is to use the VPF to understand what precision in reionization constraints the narrowband survey of LAGER will be able to find as a function of area.
\citet{Perez2021} derived conservative guidelines for when a VPF measurement is reliable, as well as minimum requirements of survey sizes to meet these guidelines. We apply these guidelines to the case of the LAGER survey below, and examine the VPF and its ability to to constrain the ionization fraction in detail in \textsection \ref{sec:GetXiFractionfromVPF}. Additionally, future reionization studies can use these guidelines and our analysis as a starting point to prepare for LAE surveys that will measure clustering.

A minimum radius for the VPF is simply the average distance between two random points, and relies on the density of the sample. Any smaller empty test spheres may not be true voids in the sample, but perhaps consequences of resolution. The maximum radius in this framework depends on the entire survey volume and the desired precision of the VPF. In order to measure $\log_{10}(P_0)$ to a given $-\alpha$ value and guarantee the level of precision of the VPF within $\pm 10^{-\alpha}$, the survey volume must be able to contain $10^{\alpha}$ independent sub-volumes of the given radius. By considering the survey size and density, we can derive the number of galaxies needed for to measure the VPF to a given precision across a given dynamic range \textit{d}. For the 2D VPF, these guidelines use survey area $A$ and surface density $\Sigma$:

\begin{equation}
    R_{\text{min}} = \sqrt{\frac{1}{\Sigma \pi}}; \quad R_{\text{max}} = \sqrt{\frac{A}{\pi 10^{\alpha}}}; \quad d = \frac{R_{\text{max}}}{R_{\text{min}}} =  \sqrt{\frac{\Sigma A}{10^{\alpha}}}; \quad N_{\text{total}} = \Sigma \times A = d^2 10^{\alpha}
\label{eq:2Dlimits}
\end{equation}


The observed number density of LAEs is  $2.86\times 10^{-5}$ Mpc$^{-3}$ across the first 4 LAGER fields \citep{Wold2022}, where each DECam field images about 3.3 deg$^2$ and the effective areas the LAEs cover is about 2.5 deg$^2$ per field. Once converted to a surface density of 17.1 LAEs deg$^{-2}$, this yields a conservative minimum radius of approximately 8.2 arcminutes. For a single LAGER field of approximately 2.55 deg$^2$, the maximum radius to measure to log$_{10}(P_0)=-1.5$ is 9.5 arcminutes. For the four-field effective area of 10.2 deg$^2$, the maximum radius to measure to the more precise log$_{10}(P_0)=-2$ is 11 arcminutes. Finally, for the full-face simulation area of 15.5 deg$^2$, the maximum radius to measure to log$_{10}(P_0)=-2$ is 13.3 arcminutes. Based on the VPF measurements in Figures \ref{fig:LcutAllFracs_LAGER}-\ref{fig:LcutOnlyRmax_Histograms_LAGER} (described completely in \textsection \ref{sec:GetXiFractionfromVPF}), values of $\alpha_{\text{2D}}>-1.5$ for the VPF easily encompass the large differences across ionized fractions and remain within our recommended guidelines. 

However, we can additionally leverage our results in this work to refine the minimum VPF radius in a more practical way. For example, one could say that the minimum radius to measure the VPF is when the measurement is statistically distinct from the VPF of an unclustered distribution. The 2D VPF of the full-face slices in Figure \ref{fig:LcutAllFracs_LAGER} begin to lie at least 3$\sigma$ away from the Poisson curve near at least 2 arcminutes (if using the standard error pictured in Figure 2) or more realistically near 5 arcminutes (if instead using just the standard deviation). Therefore, we can update our conservative lower radius limit of the full 8-field LAGER (equal in area to one of the 20 slices) VPF measurement to be 5 arcminutes.

\section{Constraining Global Neutral Fractions with the VPF from LAGER}\label{sec:GetXiFractionfromVPF}

In this section, we ask and answer: what area is needed to make precise constraints on reionization for the LAGER survey with LAE clustering? Specifically, how distinguishable are the different ionization fractions using the VPF for narrowband-detected volume limited samples as a function of survey volume? To do this, we use the conservative guidelines of \citet{Perez2021} to identify which radii will give trustworthy VPF measurements, and focus upon the behavior of the VPF at the smallest and largest trusted scales for when we pick and sample one, four, or eight LAGER-like fields.

As discussed in \textsection \ref{sec:EoRClustering_General}, the clustering of LAEs is affected in its extent and distribution by reionization. Instigating a single luminosity cut across all ionization fraction volumes mimics how narrowband-detected LAEs at different redshifts and neutral fractions show a clear drop in their number density and increase in their clustering. This closely complements analyses of the Ly$\alpha$ luminosity function as it evolves through the epoch of reionization, and will offer a constraint of the neutral fraction of the IGM under the same effect with different models and simulations, and utilizing the same data with very little modification. 

\begin{figure*}
	\begin{center}
    \includegraphics[width=\textwidth]{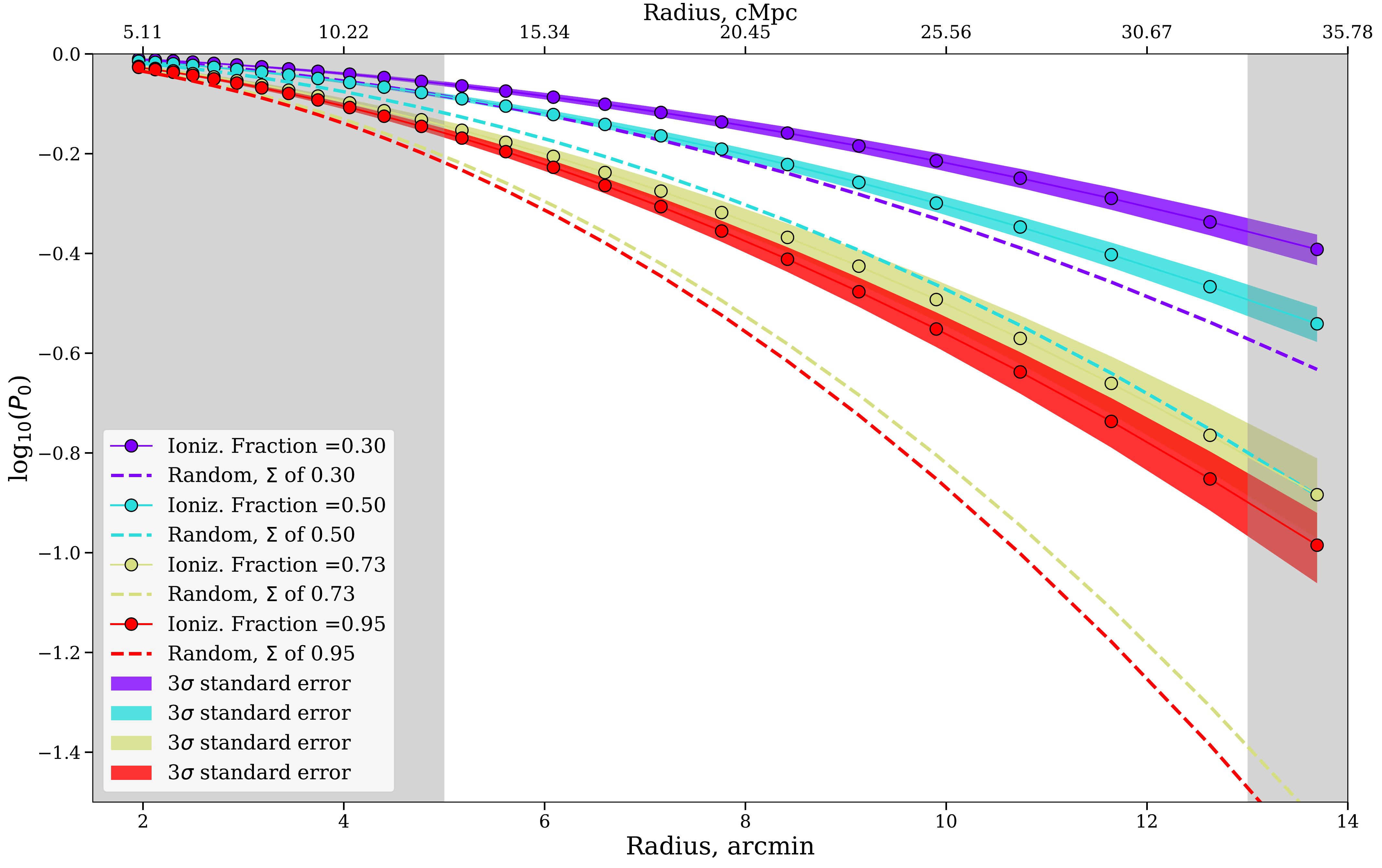}
	\caption{The 2D VPF of LAGER-like simulation slices of \citet{Jensen2014} at (30, 50, 73, 95) percent ionized hydrogen fractions in (purple, blue, yellow, red) respectively. We assume a Ly$\alpha$ luminosity cut of log$_{10}$L$_{\text{Ly}\alpha} >$ 42.9 erg s$^{-1}$, corresponding to the approximate the line flux where the observed LAGER LAEs are 50$\%$ complete. The (602x607x600) Mpc$^3$ simulations have been divided into 20 slices of (602x607x30) Mpc$^3$, corresponding to the approximate depth of the LAGER narrowband at $z_{\text{Ly}\alpha} \sim 6.9$ and effective area projected for the 8 planned LAGER fields. With these large slices, we can approximate the `true' 2D VPF for the simulated LAEs at each ionization fraction (by covering large enough areas to minimize cosmic variance), and later compare to the 2D VPF we measure with LAGER DECam-sized subslices of these simulations.
	We plot the mean 2D VPF across the 20 slices as colored circles and shade the $3\sigma$ standard error (3 standard deviation divided by the square root of the number of samples). Different ionization fractions show varying number densities and very different VPFs, both from the effect of decreasing neutral fractions and also evolution in the host dark matter halos and galaxies themselves over cosmic history. 
	\label{fig:LcutAllFracs_LAGER}
	}
	\end{center}
\end{figure*}

\begin{figure*}
	\begin{center}
    \includegraphics[width=\textwidth]{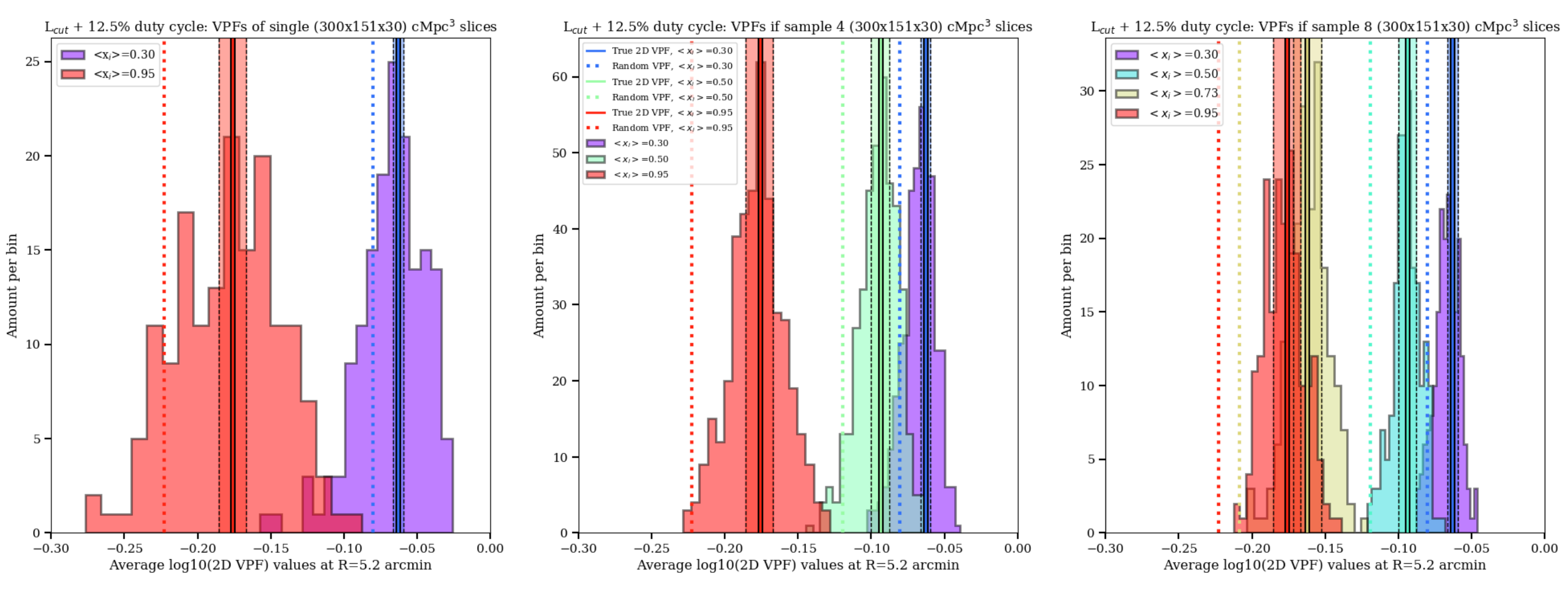}

	\caption{
	What constraints on ionization fraction can the VPF give for various survey areas of LAGER, at a small VPF scale?
	We plot the distribution of the 2D VPF measured at 5.2 arcmin (13.23 Mpc) for: all 160 individual DECam-sized subslices (left), the average when randomly choosing four subslices (center) four hundred times, and the average when choosing eight subslices (right) two hundred times. Sampling more subslices will lead to averages that are more narrowly distributed around the true average value. The solid lines are the 2D VPF of the full-face volumes for the given ionization fraction in Figure \ref{fig:LcutAllFracs_LAGER}, and serve as our approximation of the `true' VPF value at this radius. The 3$\sigma$ standard error from of the full-face 2D VPF are the colored shaded regions enclosed with thin black dotted lines.
	The VPFs at this radius for completely unclustered Poisson distribution at the relevant surface densities (averaged across the 20 whole-face slices) are approximately \{-0.08, -0.13, -0.23\} for \{0.30, 0.50, 0.73/0.95\}.
    The left panel shows that a single LAGER field could constrain between 30$\%$ or 95$\%$ ionized universe; the center shows that four averaged fields might additionally distinguish 50$\%$ ionized around -0.15$<$ log$_{10}$P$_0 <$ -0.1; and the right shows that eight averaged fields more clearly distinguish 50$\%$ ionized, and may in rare cases also separate 73$\%$ from 95$\%$ ionized fraction where -0.16 $<$ log$_{10}$P$_0 <$ -0.12. 
    \label{fig:LcutOnly_Histograms_LAGER}
    } 

	\bigskip

	\includegraphics[width=\textwidth]{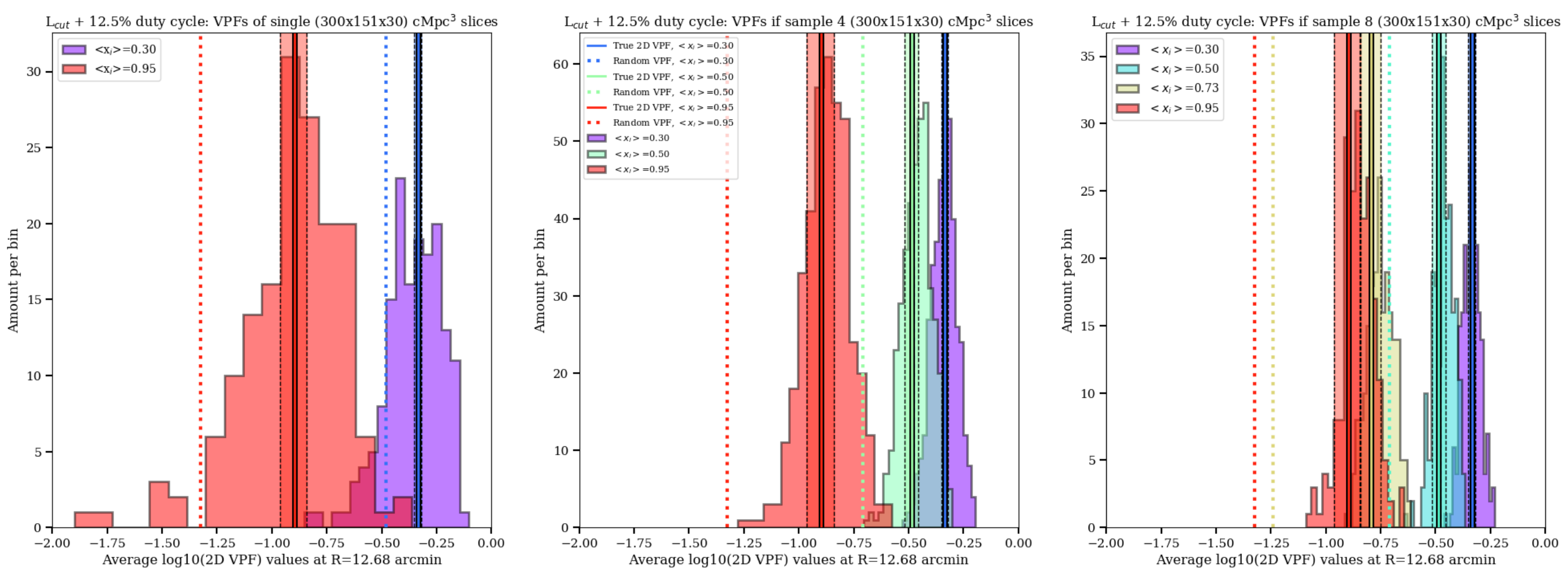}
    
    \caption{Like Figure \ref{fig:LcutOnly_Histograms_LAGER}, but measured at the maximum radius of 12.68 arcmin (32.27 cMpc). The VPFs at this radius for completely unclustered Poisson distribution at the relevant surface densities (averaged across the 20 whole-face slices) are approximately \{-0.5, -0.75, -1.25, -1.3\} for \{0.30, 0.50, 0.73, 0.95\}, and shown in colored dotted lines. The 3$\sigma$ standard error from of the full-face 2D VPF are the colored shaded regions enclosed with thin black dotted lines.
    Like Figure 3, these distributions show that the VPF measured at large scales can also constrain whether an individual LAGER field is more likely to be 30$\%$ or 95$\%$ ionized, and in some cases statistically distinguish a 50$\%$ ionization fraction under this model as well. Individual VPF measurements become less correlated the more separated they are in distance, allowing a verification of constraints at small distance scales. 
    \label{fig:LcutOnlyRmax_Histograms_LAGER}
    }    
	
	\end{center}
\end{figure*}

\begin{table*}[t] 
    	\begin{center}
    \caption{Details of the VPF distributions in Figures \ref{fig:LcutOnly_Histograms_LAGER} and \ref{fig:LcutOnlyRmax_Histograms_LAGER}. Once cutting up the \citet{Jensen2014} simulations into 160 LAGER DECam-sized slices of (301x150.5x30) Mpc$^3$, we measure VPF at the minimum and maximum scale for LAGER across all individual slices, and then when randomly picking and averaging four or eight slices. We find the distributions are well described by Gaussian functions, and we list the mean log$_{10}$VPF value and full width at half max for each distribution. We also calculate the approximate overlap between the simulations' curves for each sampling, finding what percentage of a given simulation's VPF are measured at the same value in any another simulation, and accounting for variations due to binning and the sampling.
    \label{table:VPFhistogramDetails}
    }   
	\begin{tabular}{p{2.75cm}p{1cm}p{1.5cm}p{1.25cm}p{2.5cm}p{1.5cm}p{1.25cm}p{2.5cm}} 
	\hline \hline
    \multicolumn{2}{c}{Histogram selection} & \multicolumn{3}{c}{R$_{\textrm{min}}=5.2$ arcmin}  & \multicolumn{3}{c}{R$_{\textrm{max}}=12.7$ arcmin}  \\
    \hline

        Sample size & $\langle x_i\rangle _m$ & Peak VPF & FWHM & $\%$ overlap & Peak VPF & FWHM & $\%$ overlap \\
        \hline
         All 160 single slices & 0.30 & -0.061 & 0.026 & 20$\%$ 0.30-0.95 & -0.31 & 0.12 & 40$\%$ 0.30-0.95 \\
		... & 0.95 & -0.169 & 0.038 & ... & -0.83 & 0.19 & ... \\
        Pick 4, 400 times & 0.30 & -0.067 & 0.012 & $>1\%$ 0.30-0.95 & -0.33 & 0.06 & 0$\%$ 0.30-0.95 \\
    	... & 0.50 & -0.093 & 0.014 & 80$\%$ 0.30-0.50  & -0.46 & 0.08 & 80$\%$ 0.30-0.50 \\
    	... & 0.95 & -0.169 & 0.02 & 18$\%$ 0.50-0.95 & -0.81 & 0.10 & 10$\%$ 0.50-0.95 \\
        Pick 8, 200 times & 0.30 & -0.067 & 0.0085 & 0$\%$ 0.30-0.73  & -0.33 & 0.041 & 0$\%$ 0.30-0.73 \\
        ... & 0.50 & -0.092 & 0.013  & 35$\%$ 0.30-0.50 & -0.46 & 0.057 & 50$\%$ 0.30-0.50 \\
        ... & 0.73 & -0.15 & 0.012 & 0$\%$  0.50-0.73 & -0.71 & 0.076 & 1$\%$ 0.50-0.73 \\
    	... & 0.95 & -0.17 & 0.016 & 90$\%$ 0.73-0.95 & -0.81 & 0.079 & 90$\%$ 0.73-0.95 \\
    \hline
	\end{tabular}
	\end{center}
\end{table*}

As derived in \textsection \ref{subsec:WhenuseVPF}, a \textit{single} DECam field of 3.3 deg$^2$ at the observed LAE surface density of 4.74$\times 10^{-3}$ arcmin$^{-2}$ will yield the conservative minimum and maximum radii limits of about 8 and 10 arcminutes (respectively). In recognizing that real galaxies will be clustered enough to be statistically distinct from random at smaller radii, we choose to focus on the 2D VPF measured at $R=5.2$ arcminutes. At this radius, the mean VPF value of the 20 full-face slices is at least 3 standard deviations away from the predicted value for a random distribution for all the simulations. In comoving units, this corresponds to 13.3 cMpc, within our 25 radii spaced logarithmically between 5 and 35 Mpc. To contrast this, we examine the VPF at 12.68 arcminutes (32.27 cMpc), near the largest radius that our conservative guidelines suggest for the fully planned 8 LAGER fields (assuming a similar yield of effective area as seen in \citealt{Wold2022}, approximately 15.5 deg$^2$).

In Figure \ref{fig:LcutAllFracs_LAGER}, we show the 2D VPF under a single luminosity cut for the 20 whole-face slices of (602x607x30) cMpc$^{3}$. This mimics the `true' value and inherent variance of the 2D VPF for 
narrowband-selected LAEs at $z=6.9$, given the simulation assumptions and models. The colored circles are the mean VPF across the 20 slices, the shading is the 3$\sigma$ standard error (where 1$\sigma$ standard error is one standard deviation divided by the square root of the number of slices), and the dashed lines are the theoretical 2D VPF for an unclustered distribution at the surface density of the given simulation. Each ionization fraction simulation is strongly clustered compared to its random curve; and promisingly for our analysis, each ionization fraction's 2D VPF curves is statistically distinct from the others. As we will find when examining the spread of the 2D VPF across individual DECam-sized slices, the large separations allows for us to constrain mostly ionized vs. mostly neutral given the \citet{Jensen2014} model assumptions with the four currently observed LAGER fields.

To probe how survey area affects the constraints that LAGER might give for reionization using the 2D VPF of LAEs, we leverage the size of the \citet{Jensen2014} simulations to examine the behavior of the 2D VPF across 160 subslices similar to individual DECam imagings. In Figures \ref{fig:LcutOnly_Histograms_LAGER} and \ref{fig:LcutOnlyRmax_Histograms_LAGER}, we plot the measured 2D VPFs under one Ly$\alpha$ luminosity cut (essentially one flux limit in our narrowband).The leftmost subplot shows the distribution of 2D VPFs for all 160 (300x150x30) Mpc$^3$ individual subslices (left); the center shows the distribution when we randomly "pick 4" subslices and average them four hundred times; and the left shows the distribution when we randomly "pick 8" subslices and average them two hundred times. The "pick 4" and "pick 8" strategy mimics how the LAGER team will approach the clustering measurements of the four current and eight total planned fields. In Table \ref{table:VPFhistogramDetails}, we list for Figures \ref{fig:LcutOnly_Histograms_LAGER} and \ref{fig:LcutOnlyRmax_Histograms_LAGER} the approximate central log$_{10}$VPF value of each distribution, the full width at half maximum, and approximate percent overlap of each simulation's distribution with its neighbors'.

When assessing the distribution of the 2D VPF for individual subslices, the 30$\%$ and 95$\%$ simulations are quite distinct. As seen in Figure \ref{fig:LcutOnly_Histograms_LAGER} and quantified in Table \ref{table:VPFhistogramDetails}, the simulations overlap at their only tails around log$_{10}$P$_0 \approx -0.13$. LAEs in a given LAGER field are more likely to be mostly ionized if their log$_{10}$P$_0$ is measured to be less than -0.13, and mostly neutral if their log$_{10}$P$_0$ is greater than -0.13. When we randomly sample and average four fields (aka. "pick 4"), the center panels of Figures \ref{fig:LcutOnly_Histograms_LAGER} and \ref{fig:LcutOnlyRmax_Histograms_LAGER} show how the 30 and 95 $\%$ ionized fraction simulations completely separate with no overlap, allowing one to confidently determine that the $z=6.9$ universe is consistent with being either almost entirely ionized or mostly neutral with the VPF of four LAGER fields. Additionally, the 50$\%$ simulation fills much of the gap between them, even allowing one to tell apart 30 vs. 50$\%$ ionized fractions near log$_{10}$P$_0=-0.1$ ($-0.5$ in Figure \ref{fig:LcutOnlyRmax_Histograms_LAGER}). With eight fields sampled (aka. "pick 8"), as pictured on the right side of Figure \ref{fig:LcutOnly_Histograms_LAGER}, the distributions will become narrower and allow more specific ionization fraction measurements by the \citet{Jensen2013} models, differentiating even 73$\%$ and 95$\%$ ionized fractions for rare samplings. For example, near log$_{10}$VPF=-0.14 in Figure \ref{fig:LcutOnly_Histograms_LAGER} or log$_{10}$VPF=-0.6 in Figure \ref{fig:LcutOnlyRmax_Histograms_LAGER}, the 95$\%$ distribution nearly never samples but the 73$\%$ distribution often does.

In Figure \ref{fig:LcutOnlyRmax_Histograms_LAGER}, we examine the VPF distributions at 12.68 arcmin (32.27 cMpc), the largest distance scale our covered area allows us to measure. This corresponds to the 13 arcminutes allowed by the projected 15.5 deg$^2$ for the full 8 fields planned for LAGER for a VPF sensitivity of log$_{10}(P_0)=-\alpha=-2$. The correlations between individual VPF measurements will decrease the more separated they are in distance scales, and VPF measurements are least correlated at small scales and most correlated at the largest scales (\citealt{Gangolli2021} and Gangolli, private communication). This allows us to confirm the constraint on the ionized hydrogen fraction with the VPF from the smallest distance scale, while minimizing the correlation between the clustering measurements.

As with the measurement at our smallest distance scale, at this largest distance scale, the separation between 30$\%$ and 95$\%$ ionized is significant with even a single DECam field. The 30$\%$ and 95$\%$ ionized simulations can be further distinguished from 50$\%$ with 4 fields, and in rare cases also 73$\%$ ionized with all 8 planned fields of LAGER. We stress again that we have somewhat pessimistically assumed the next 4 fields imaged by LAGER will cover the same effective area of the completed 4 fields. If the 4 fields currently undergoing imaging and analysis all have effective areas of more than 3 deg$^2$ (as did WIDE12), the increase in both the number of LAEs detected and covered area will mean the VPF can be measured more precisely to larger dynamic range, with clearer understanding of cosmic variance, and further constrain the IGM ionization fraction at $z=6.9$.

Finally, let us further focus on the possibility of constraining between the most ionized fractions, as various studies attempting to constrain late stage reionization have explored (e.g.\ \citealt{Gangolli2021}). We predict that the 8 planned fields of LAGER will in rare circumstances distinguish 73$\%$ from 95$\%$ ionized. This comes from examining the third panels of Figures \ref{fig:LcutOnly_Histograms_LAGER} and \ref{fig:LcutOnlyRmax_Histograms_LAGER}: the 73$\%$  from 95$\%$ distributions overlap significantly (as Table \ref{table:VPFhistogramDetails} notes pessimistically, about 90$\%$ of these distributions overlap in any amount). Therefore, the 8-field VPF measurement would need to lie at the far edges of either distribution (i.e. be a particularly clustered 73$\%$ value, or particularly \textit{un}clustered 95$\%$ value) to distinguish the fractions. We do see that the full uninterrupted `full-face' $\sim$24 deg$^2$ measurement of the VPF separates the 73$\%$ and 95$\%$ distributions to 3$\sigma$. Assuming that any unknown systematic features in the observed LAGER sample do not affect their clustering signal, we can ponder: how many DECam fields' worth of imagining are needed to reach a similarly narrow distribution? As Figures \ref{fig:LcutOnly_Histograms_LAGER} and \ref{fig:LcutOnlyRmax_Histograms_LAGER} show, sampling 8 fields begins to narrow the distributions enough to occasionally distinguish; therefore, perhaps 10 or more fields of LAGER, for at least 30 deg$^2$ total, could better statistically distinguish the 73 and 95 $\%$ ionized.
This is consistent with \citet{Gangolli2021}'s analysis with the $z=5.7$ LAE VPF in e.g.\ Figure 6, where the VPF ``1$\sigma$ dipersion from cosmic variance in the VPF (assumed) a futuristic... SILVERRUSH-like survey with roughly double the size" (approximately 40 deg$^2$). 
Under both their results and ours, \textit{and barring unknown systematics that may emerge once shifting to observations}, the VPF of a larger LAGER survey could therefore further refine \citet{Wold2022}'s constraint of $\langle x_{i, \textrm{LAGER}} \rangle > 0.67$ or $\langle x_{\textrm{HI, LAGER}} \rangle < 0.33$.

\section{Summary and Conclusions} \label{sec:Conclusion}

We leverage the uncommon VPF statistic to quantify how precise constraints on reionization from clustering in the Lyman-$\alpha$ Galaxies in the Epoch of Reionization (LAGER) narrowband survey may be. Specifically, we probe how distinguishable different ionization fractions are when using the VPF for narrowband-detected volume limited samples as a function of survey volume.
We use the simulations of LAEs throughout reionization from \citet{Jensen2014}, and leverage the large (600 Mpc)$^3$ volumes to statistically probe what the VPF of LAGER might measure for one, four, or eight DECam imagings of LAEs at $z=6.9$. We apply a Ly$\alpha$ line luminosity cut of log$_{10}$L$_{\text{Ly}\alpha}>$ 42.9 erg s$^{-1}$, found to be the approximate 50$\%$ completeness limit for the observed LAEs of \citet{Wold2022}.
Finally, we create volumes of LAEs within an IGM whose hydrogen is 30, 50, 73, or 95$\%$ ionized, slicing each to the width that the LAGER N964 filter spans at $z=6.9$, approximately 30 cMpc.


We measure the VPF on the 20 whole-face (602x607x30) Mpc$^3$ slices as the `true' VPF with minimal cosmic variance, given the \citet{Jensen2013} models and the LAGER flux detection limits, in Figure \ref{fig:LcutAllFracs_LAGER}. We then create slices of (301x150.5x30) Mpc$^3$ to serve as samples of what a single DECam imaging might measure, based on the effective areas across the four fully-imaged LAGER fields analyzed in \citet{Wold2022}. To understand what constraints for reionization the cumulative LAGER survey might give with the VPF, we randomly sample four (eight) subslices four (two) hundred times, creating distributions of what the VPF measurement might look like for the four already analyzed LAGER fields and with the additional four fields in progress (Table \ref{table:VPFhistogramDetails}). We examine the VPF distributions near the smallest distance scale given the observed LAE surface density (near 5 arcmin), and the maximum radius implied by the full 8 field LAGER effective area (near 13.5 arcmin).

We note that all the \citet{Jensen2014} simulations' morphology for reionization comes from the single ionization scenario that was simulated, and that the shapes of neutral and ionized regions will change with other assumptions for the sources and sinks of Ly$\alpha$ emission. Our results are model-dependent upon the single scenario run, and offer an additional comparison to models used for other Ly$\alpha$ studies. Additionally, the large size of the simulations allows us to answer how much volume is needed to constrain ionization fraction using the VPF, especially for the unique geometry of LAGER. Future work, for example, will extend this type of analysis to determining what constraints on models for the pacing of ionization the Roman Space Telescope will be able to give for deep and wide surveys of LAEs between $7.2<z<14$ (Perez et al., in prep).

Under the \citet{Jensen2013} models and with the \citet{Jensen2014} simulations, we find that even a single DECam field might be able to discriminate between mostly neutral or mostly ionized based on the VPF distributions in Figures \ref{fig:LcutOnly_Histograms_LAGER} and \ref{fig:LcutOnlyRmax_Histograms_LAGER}. Sampling and averaging four fields allows the distinction between 30, 50, and 95 percent ionized, and utilizing eight or more fields can distinguish between 73 and 95 percent ionized in outlying circumstances. An expanded LAGER survey of ten or more fields, with 30+ deg$^2$ would be expected to more confidently distinguish 73 and 95 percent ionized (within $3\sigma$), and perhaps better refine \citet{Wold2022}'s constraint of $\langle x_{i, \textrm{LAGER}} \rangle > 0.67$ or $\langle x_{\textrm{HI, LAGER}} \rangle < 0.33$. Finally, as proposed in \citet{McQuinn2007}, the VPF of a fully ionized sample under the same luminosity selection (e.g. narrowband LAEs at z=5.7 or below) can be compared to the VPF of the LAGER $z=6.9$ LAEs to further constrain reionization with \citet{Jensen2013} model. 
Future work in the LAGER collaboration will examine the VPF and other clustering statistics across the LAEs of all completed and ongoing fields, and be informed by the precision and constraints predicted by our work.

\begin{acknowledgments}
We thank Hannes Jensen and Garrelt Mellema for sharing the simulations that this work analyzed. We also thank Garrelt Mellema for comments offered that improved the manuscript. We thank the larger LAGER collaboration for their support of this work. We thank the anonymous reviewer for their comments that improved the clarity and context of this manuscript. Finally, we are grateful for the continual support from Isak Wold as we refined our experiment setup using \citet{Wold2022}. This work was supported by NSF grant AST-1518057, as well as NASA support to A.S.U. under contract NNG16PJ33C, “Studying Cosmic Dawn with WFIRST.” The Cosmic Dawn Center (DAWN) is funded by the Danish National Research Foundation under grant No. 140. Part of this work was carried out on the \textit{Saguaro} supercomputing cluster operated by the Fulton School of Engineering at Arizona State University, and also made use of NASA's Astrophysics Data System Bibliographic Services. Our figures were made with \textit{Matplotlib} \citep{Matplotlib}, and the bulk of our calculations used \textit{Numpy} \citep{Numpy}, \textit{Scipy} \citep{Scipy}, and the \citet{NedWright} cosmology calculator.

\end{acknowledgments}

\newpage 

\appendix


As detailed in \textsection \ref{subsec:LAGERsetup}, we select mock LAGER LAEs with a single cut on the Ly$\alpha$ line luminosity of log$_{10}$ L$_{\text{Ly}\alpha}>42.9$ erg s$^{-1}$. This corresponds with the approximate cutoff luminosity that yields a 50$\%$ complete sample across the 4 observed fields of LAGER in \citet{Wold2022}. We have confirmed that this straightforward approach yields functionally identical clustering results and conclusions when implementing the detailed luminosity-dependent completeness selection of the observed LAGER LAEs. We detail this process and provide its version of Figure \ref{fig:LcutAllFracs_LAGER} here.

\begin{figure*}[b]
	\begin{center}
    \includegraphics[width=\textwidth]{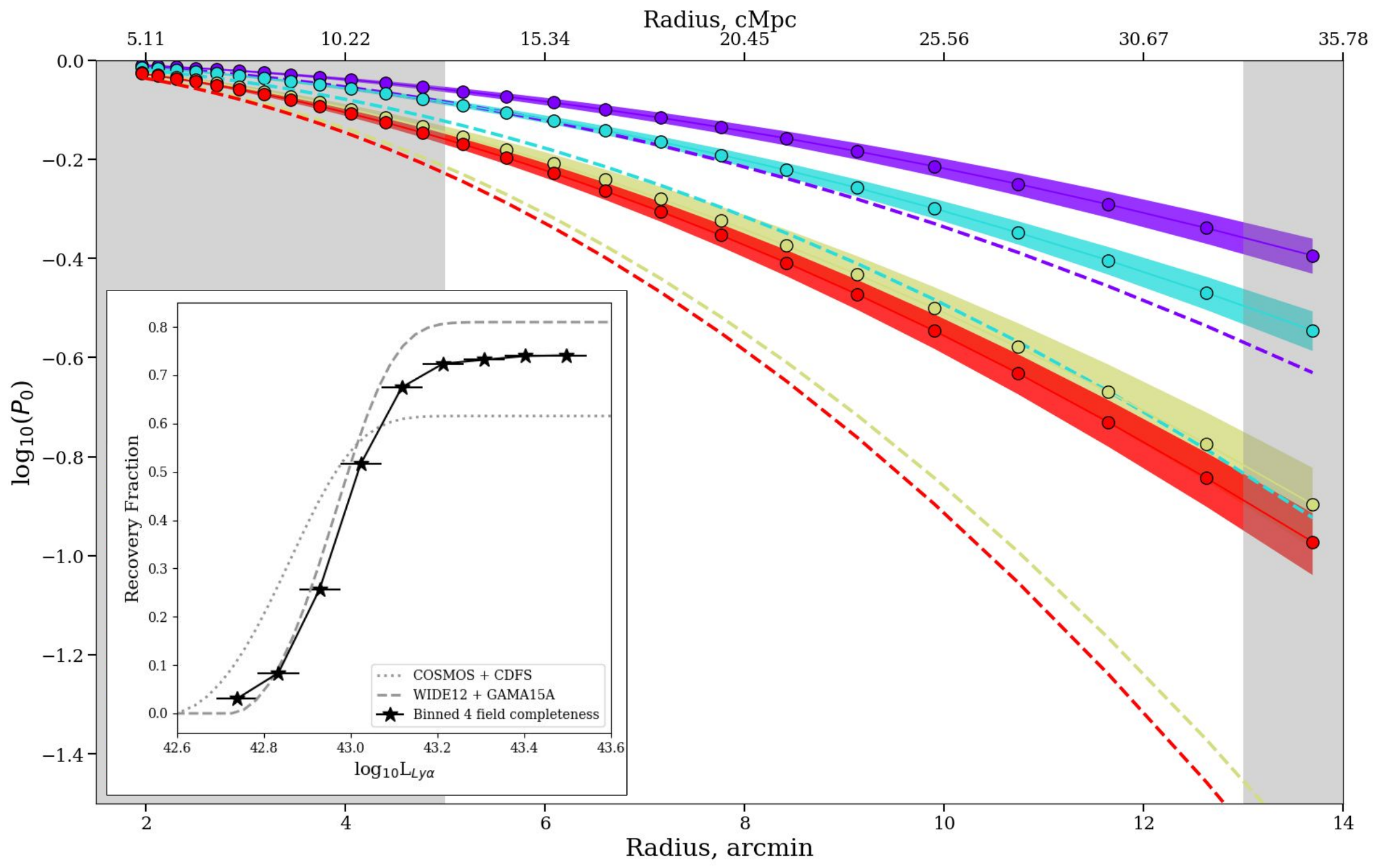}
	\caption{Like Figure \ref{fig:LcutAllFracs_LAGER}, but when instead applying the luminosity-completeness relationship derived by \citet{Wold2022} to create the 4 field LAGER Ly$\alpha$ luminosity function at $z=6.9$ (window figure). Comparing the VPF curves here to those in Figure \ref{fig:LcutAllFracs_LAGER} confirms that using a straightforward luminosity cut yields functionally identical VPF clustering results for LAGER.
	\label{fig:LAGERcompleteness}
	}
	\end{center}
\end{figure*}

We transform the narrowband magnitude vs.\ recovered LAE fraction relationships of the final LAE selection for the WIDE12 and GAMA15A fields of LAGER (Fig.\ 6 of \citealt{Wold2022}) into a function of Ly$\alpha$ line luminosity vs.\ recovered fraction, shown in a window plot within Figure \ref{fig:LAGERcompleteness}. This luminosity-completeness function is calculated for each bin in the LAGER 4 field luminosity function (Fig.\ 10 in \citealt{Wold2022}). For our analysis, the completeness of a particular bin of Ly$\alpha$ luminosity is treated as the probability of a LAE being detected. We apply this luminosity-dependent probability of selection across all of the \citet{Jensen2014} LAE catalogs, as well as a 40$\%$ duty cycle (corresponding roughly to the difference between the LAGER LF and the $z=5.7$ LAE LF of \citealt{Ouchi2010} that the catalogs are most consistent with, and again bringing the most ionized catalogs' number densities into agreement with LAGER). We re-plot the 2D VPF curve across the catalogs as otherwise described in Figure \ref{fig:LAGERcompleteness}. Figure \ref{fig:LAGERcompleteness} is nearly identical to Figure \ref{fig:LcutAllFracs_LAGER}, finding only slight differences in the final densities of each catalog, and slightly more overlap in the wings of the 73$\%$ and 95$\%$ ionized catalogs. Therefore, our conclusions in \textsection \ref{sec:GetXiFractionfromVPF} are expected to hold for the observed LAE clustering under the current LAGER selection.

\clearpage
\bibliography{sample63}{}
\bibliographystyle{aasjournal}

\end{document}